\documentclass[letterpaper,twocolumn,10pt]{article}
\usepackage{usenix,epsfig,endnotes}
\usepackage{pifont}
\begin{document}

\date{Thursday, February 13}

\title{\Large \bf Estimating verification time}

\author{
{\rm Pablo Gonz\'alez de Aledo}\\
Universidad de Cantabria
}

\maketitle

\pagestyle{empty}

This essay is divided in four parts: In section (I) I explain why I think detecting hot-spots in verification is complicated and in particular, more complicated than detection when developing crude software. In section (II) I introduce the factors I think mostly affect performance in verification. In section (III) I propose a method to find functions that are most promising to be optimized. Finally, in section (IV) I draw some conclusions and discuss pros/cons of the proposed solution.

\section{Why detecting hot spots in verification is difficult}

A generic framework for automated program verification through symbolic execution comprises the following steps:

\begin{enumerate}
\item The code is parsed and transformed into an intermediate representation for instrumentation.
\item This intermediate code is instrumented for verification and compiled again to make it executable.
\item While running, the program forks on every branch instruction, and calls a solver that decides which branches can be executed.
\item The annotated code and the solver run over an operating system, which schedules the different segments of code, commonly over an SMP processor, but sometimes over a distributed set of processing nodes, each of them with its own memory, scheduler, branch-prediction and memory management directives...
\end{enumerate}

Many of these transformations and processes --particularly the transformations of step \ding{172}, the forking scheme and the constraint solver of step \ding{174}-- are guided by heuristics that have been tuned and optimized during years, so it is impossible in practice to obtain a formal method to derive execution time directly from the high-level description of the function under test. 

\begin{figure*}[!]
\centering
\includegraphics[width=\textwidth]{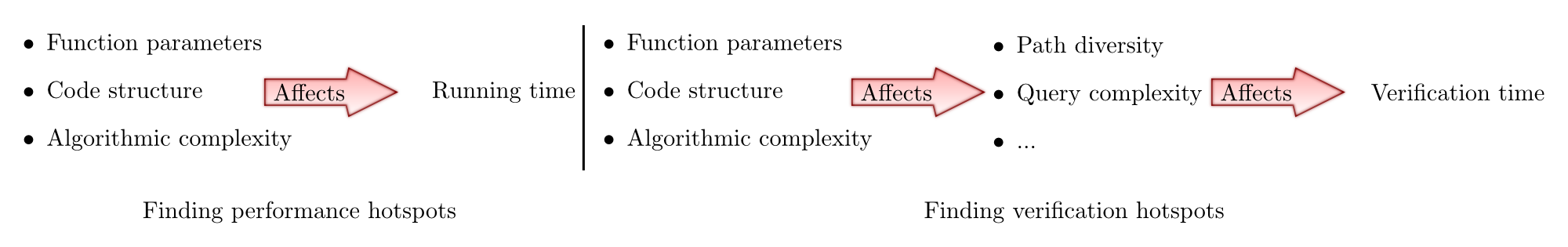}
\caption{Hot-spot finding when programming and when verifying.}
\label{affects}
\end{figure*}

Also, in the particular case of verification, the function parameters, the structure of the code or the algorithmic complexity does not directly affect performance, but instead, affect it through hidden parameters like path diversity and query complexity (Figure \ref{affects}). The challenge in this case is therefore twofold; first, we have to describe the effect of visible characteristics of the code into these hidden parameters and then, the effect of intermediate factors into performance.

\section{What affects performance in verification}

Performance of symbolic execution is dominated by the following factors (in order of importance):
\begin{itemize}
\item Paths to explore (path diversity), greatly affected by the number of loops with variable number of iterations.
\item Complexity of branch-conditions that are symbolic; number of variables, depth and kind of query (linear or not).
\item Memory accesses.
\item Arithmetic operations.
\end{itemize}

When considering how particular factors affect verification time, we have to take account of the compiler transformations, and measure the above factors over optimized code. Among the things that the compiler can do to make code more amenable for verification, are : algebraic simplifications, removal of redundant operations and constant propagation; loop unrolling; discard CPU-specific optimizations; aggressively inline functions... (\cite{overify})

\section{A method to find functions that are more promising to be optimized.}

As explained in the previous sections, it is difficult and probably undecidable to derive verification time from a set of factors that can be observed in the code. This is why the best way to confront the problem is a macro-modelling approach \cite{Automated}. In macro-modelling, we map a piece of code to a model that relates some output dependent of this code with various parameters that can be obtained, observed or calculated easily from the high-level description. Important aspects of this technique are the selection of the formula (linear or non-linear), the selection of input parameters, and the fitting process.

In particular, I will follow a white-box analysis and black-box measurement philosophy.
\begin{itemize}
\item White-box analysis means that we have access to the variables that affect verification time, we can modify them with little constraints and devise experiments that illustrate the effects of these changes.
\item Black-box measurement means that we measure the verification time at the end of the previously commented chain, when all the transformation and heuristics are considered. Black-box measurement enable the time models to be estimated with little knowledge of the steps involved in optimization and constraints solving.
\end{itemize}

Challenges in the practical application of this approach include:

\begin{itemize}
\item Identifying the parameters and the template function of the macro-model will require significant manual effort. To choose a correct set of parameters and still keep generality to handle the wide range of behaviors of C-standard library is quite challenging.
\item Designing a proper set of experiments to measure the effect of parameters into verification performance.
\item Developing an accurate model that relates verification performance with the code and input data.
\item Estimating the contribution of each parameter to the verification time.
\item Isolating errors inherent to measures.
\item Obtaining confidence intervals for inputs and outputs.
\item Validating the model.
\end{itemize}

In order to enable human intervention to help automated methods, I have divided the methodology in 5 stages:

\subsection{Parameter identification}

In the first step, we identify the essential set of factors that may be correlated with performance. This is arguably the most difficult step in the process and the one that most affects both the accuracy and practicality of the approach. In general, any field nested in the input structures of the function and any local variable inside the function under study can affect verification time. However, the number of fields in structures passed to stdlib functions generates a lot of potential parameter candidates, even for simple functions, and the design space of all those variables is too big to be exhaustively analyzed. Therefore, a good automated tool must include a step to characterize which of these parameters are truly relevant for the problem, and which can be omitted. In some cases, human insight can help selecting those parameters, but the effort of manually performing this task often hinders the application of these methods in practice, so in order to keep the method practical, human intervention should be allowed but reduced as much as possible.

I have chosen the following parameters as candidates for the next step:

\begin{enumerate}
\item Cyclomatic complexity of the function.
\item An array indicating for each symbolic loop inside the function, how many instructions are inside this loop.
\item An array indicating for each symbolic branch inside the function, if the query that is evaluated is linear or not.
\item An array indicating for each parameter, if it is symbolic or constant.
\item The sizes of all arrays passed to the function, either directly, or through the use of pointers. This size is particularly difficult to be obtained during run-time, as array boundaries are not maintained when an array is passed to a function. Serialization can be used to obtain a sequential description of the array \cite{Automated}, but requires manual annotation of the code. Another option is to omit size of arrays in the approach, under the empirical observation that most functions of the standard library use a separate parameter to identify this size.
\item The value of all scalar values passed to the function.
\end{enumerate}

The objective of choosing this set of factors is to approximate the effect of code structure and function parameters in path diversity and query complexity. The benefit of choosing them is that \ding{172} and \ding{173} can be obtained statically from the code, and the others does only need a light instrumentation. We still need to call a solver to measure the verification time (step 2), but the use of techniques from experimental design can reduce the number of measures we need to perform.

\subsection{Experimental design and data collection}

Second step of the methodology is concerned with obtaining the data needed to fit the macro-model for the verification time. When a minimum or close-to-minimum set of parameters have been identified, we need to define characterization experiments in order to isolate the effects of variables obtained in step 1 in verification time. "Experimental design" is a vast field that can not be easily described in this essay (and that have been largely ignored by software developers). Although references are very old, and are hard to fully automate, \cite{Art} and \cite{Design} explain ways to obtain the maximum amount of information about performance running the minimum number of experiments. The theory explained in these references shows how to construct experiments able to estimate the effects of the different parameters on the verification time as efficiently as possible.

Once the experiments have been designed, we need to instrument the code to obtain the data needed to fit the macro-models. In \cite{Automated} authors propose a method drawn from data-structure serialization as a novel technique to perform data collection, which can be applied to the particular case of verification.

\subsection{Macro-model fitting}

In the forth step, collected data are used to fit a macro-model of the functions: once the instrumented source-code has been run and data have been obtained from the experiments, we can fit the model to the data. There exists two main approaches for macro-model fitting: linear models can be easily fitted by least squares, but do not effectively capture complex inter-dependences between parameters. Symbolic regression \cite{Automated} relaxes the restriction of a linear model and therefore can construct general models for functions with highly related parameter dependencies. However, it commonly uses a hill-climbing or genetic algorithm which can fall in local-optima, obtaining sub-optimal models.

\subsection{Accuracy assessment}

The last step of the methodology is to assess the accuracy of the model: once we have obtained the macro-model it is important to evaluate the error it produces. This error provides a confidence interval of the predicted value when the model is used in subsequent estimations. Even when this confidence interval is statistical and does not consider corner-cases, it can be of practical use to justify the applicability of the model. "Experimental Design" also explains how to estimate the error of the model. We must take special care of making the set of experiments used to assess the model different than the one used to fit the data.

\section{Conclusions}

I have presented a possible method to estimate verification time of functions in the C standard library that can be used to detect which of them are responsible for a large time in the programs we verify. The benefits of the approach are:

\begin{itemize}
\item It considers many of the heuristics involved in program transformation and verification.
\item Once the method is developed, adapting it to new compilers and solvers is easy and straightforward.
\item Once the models are fitted, we do not need to actually verify the code under test, so estimating verification time is fast.
\item Statistical confidence intervals for the results can be obtained.
\end{itemize}

The cons of the proposed approach are mostly related with the selection of input parameters to the model.

\begin{itemize}
\item The use of dynamically created data structures and memory allocation complicates the identification of relevant macromodel parameters.
\item Manual selection of a function template to fit the macromodel often requires a profound understanding of the function, which may be very difficult in this scenario. Symbolic regression can help in this step, but some human intervention is still needed.
\end{itemize}

Another important limitation of the method is that the macro-model will only be applicable when the function is called with parameters that follow the same statistical distributions as the ones used to train the model, and this depends on what we are interested in analysing. For example, if we are just interested in obtaining a good coverage for a function, we will not put much effort in training the model for particular corner-cases that are seldom executed. On the other hand, in some scenarios of verification we are particularly interested in finding those corner-cases, so the model will have poor accuracy for these test-cases.

{\footnotesize \bibliographystyle{acm}

\begin{thebibliography}{9999999} 
	\bibitem{overify}
		J.Wagner, V.Kuznetsov, G.Candea,
		``-OVERIFY: Optimizing Programs for Fast Verification''
		HotOS'13 Proceedings of the 14th USENIX conference on Hot Topics in Operating Systems
	\bibitem{Automated}
		A. Muttreja, A. Raghunathan, S. Ravi, N. K. Jha,
		``Automated Energy/Performance Macromodeling of Embedded Software''
		IEEE Transactions on Computer-Aided Design of Integrated Circuits and Systems, vol. 26, no. 3, March 2007
	\bibitem{Art}
		R.Jain, 
		``Art of Computer Systems Performance Analysis Techniques For Experimental Design''
	\bibitem{Design}
		M. Schatzoff,
		``Design of Experiments in Computer Performance''
		IBM Res.Develop.vol. 25 no. 6 November 1981


\end{thebibliography}

\end{document}